\documentclass[aps,prd,twocolumn,superscriptaddress]{revtex4-2}

\usepackage{graphicx}
\usepackage{amsmath}
\usepackage{amssymb}
\usepackage{CJK}
\usepackage[colorlinks,citecolor=blue,linkcolor=blue,urlcolor=blue,anchorcolor=blue]{hyperref}
\allowdisplaybreaks
\newcommand\dx{{\rm d}}
\newcommand\p{\partial}
\newcommand\Lf{\mathcal{L}_\textsc{f}}
\newcommand\Lm{\mathcal{L}_\textrm{m}}

\newcommand\AaA{Astron. Astrophys.}
\newcommand\ApJ{Astrophys. J.}

\newcommand\CQG{Classical Quantum Gravity}
\newcommand\JCAP{J. Cosmol. Astropart. Phys.}

\newcommand\PLB{Phys. Lett. B}

\newcommand\PRD{Phys. Rev. D}
\newcommand\PRL{Phys. Rev. Lett.}

\newcommand\RMP{Rev. Mod. Phys.}

\newcommand\EurPhysJC{Eur. Phys. J. C}

\newcommand\FoundPhys{Found. Phys.}

\newcommand\IntJGeomMethodsModPhys{Int. J. Geom. Methods Mod. Phys.}
\newcommand\IntJModPhysD{Int. J. Mod. Phys. D}

\newcommand\PhysRep{Phys. Rep.}
\newcommand\PhysRev{Phys. Rev.}

\newcommand\PubAstronSocAust{Pub. Astron. Soc. Aust.}

\newcommand\SovPhysJETP{Sov. Phys. JETP}

\begin{document}

\begin{CJK*}{UTF8}{gbsn}
\title{Gravitation with modified fluid Lagrangian: Variational principle \\ and an early dark energy model}
\author{S. X. Tian (田树旬)}
\email[]{tshuxun@bnu.edu.cn}
\affiliation{Department of Astronomy, Beijing Normal University, Beijing 100875, China}
\author{Zong-Hong Zhu (朱宗宏)}
\email[]{zhuzh@bnu.edu.cn}
\affiliation{Department of Astronomy, Beijing Normal University, Beijing 100875, China}
\affiliation{School of Physics and Technology, Wuhan University, Wuhan 430072, China}
\date{\today}
\begin{abstract}
  Variational principle is the main approach to obtain complete and self-consistent field equations in gravitational theories. This method works well in pure field cases such as $f(R)$ and Horndeski gravities. However, debates exist in the literature over the modification of perfect fluid. This paper aims to clarify this issue. For a wide class of modified fluid Lagrangian, we show that the variational principle is unable to give complete field equations. One  additional equation is required for completeness. Adopting the local energy conservation equation gives the modified fluid a good thermodynamic interpretation. Our result is the first modified fluid theory that can incorporate energy conservation. As an application of this framework, we propose a specific modified fluid model to realize early dark energy triggered by cosmic radiation-matter transition. This model naturally explains why early dark energy occurs around matter-radiation equality and is useful in erasing the Hubble tension.
\end{abstract}
\maketitle
\end{CJK*}

\section{Introduction}
Generally speaking, modified gravities belong to classical field theory, in which the variational principle is an important tool to derive the field equations \cite{Capozziello2011.PhysRep.509.167,Clifton2012.PhysRep.513.1}. Fluid is an important source of gravity that describes the Universe, galaxies and stars \cite{Misner1973.book}. The equations of fluid motion are generally given by microscopic particle physics, not by the variational principle. In gravitational theories, the variational principle of general fluid is still controversial, which hinders progress in modifying gravity from the fluid side. \citet{Taub1954.PhysRev.94.1468} first constructed the Lagrangian of perfect fluid, and later \citet{Schutz1970.PRD.2.2762} gave a different but also reasonable result. \citet{Gonner1976.NotEngJ.31.1451,Gonner1984.FoundPhys.14.865} first discussed the gravitational theory with nonminimal coupling between spacetime and fluid. Two such theories that have been widely discussed recently are $f(R,\Lm)$ gravity \cite{Harko2008.PLB.669.376,Harko2010.PRD.81.044021,Harko2010.EurPhysJC.70.373,Minazzoli2012.PRD.86.087502,Wang2012.CQG.29.215016} and $f(R,T)$ gravity \cite{Harko2011.PRD.84.024020,Sharif2012.JCAP.03.028,Barrientos2018.PRD.97.104041}. A comment on the $f(R,T)$ gravity says that the pure fluid part $f(T)$ has no physical significance and the resulting theory is exactly perfect fluid \cite{Fisher2019.PRD.100.064059,Fisher2020.PRD.101.108502}. \citet{Harko2020.PRD.101.108501} refute this comment. In addition, energy is generally not conserved in $f(R,\Lm)$ and $f(R,T)$ theories. Gravitational particle creation process is needed to explain the corresponding thermodynamics \cite{Harko2014.PRD.90.044067,Pinto2022.PRD.106.044043}. Is there a way to generalize the perfect fluid that preserves energy conservation? If such a theory exists, then it can be consistent with conventional thermodynamics, which makes the theory more attractive. The debate on the $f(R,T)$ gravity and the energy conservation issue are the first two motivations for this paper.

The third motivation is an early dark energy (EDE) model we proposed in \cite{Tian2021.PRD.103.043518}. The EDE present at matter-radiation equality (redshift $\sim3400$) can be used to erase the Hubble tension \cite{Agrawal2019,Poulin2019.PRL.122.221301,Smith2021.PRD.103.123542,Hill2022.PRD.105.123536,Smith2022.PRD.106.043526, DiValentino2021.CQG.38.153001,Kamionkowski2022.arXiv.2211.04492}. However, a coincidence problem arises in the scenario --- why the energy scale of EDE is in coincidence with that of matter-radiation equality when their underlying physics seems unrelated \cite{Sakstein2020.PRL.124.161301}. Sakstein \textit{et~al.} \cite{Sakstein2020.PRL.124.161301,CarrilloGonzalez2021.JCAP.04.063} proposed a solution to this coincidence problem based on neutrino physics. Their starting point is that the neutrino mass is close to $1\,{\rm eV}/c^2$, which is exactly the energy (temperature) scale of matter-radiation equality. Using such neutrino to trigger the EDE could explain the coincidence. In \cite{Tian2021.PRD.103.043518}, we proposed a new idea that EDE may be triggered by radiation-matter transition to solve the coincidence problem. We discussed that $k$-essence \cite{ArmendarizPicon2000.PRL.85.4438} is unable to realize a viable model, and nonminimal coupling between spacetime and matter may be required. Analysis of this possibility requires a complete framework for gravitational theories with modified fluid. In this paper, we will propose a much more simple purely fluid model to realize the desired EDE.

This paper is organized as follows. Section \ref{sec:02} presents the general framework of our approach to modify fluid and a demonstration in cosmology. We emphasize that we do not consider the nonminimal coupling of spacetime geometry and fluid matter in this paper. Section \ref{sec:03} discusses the similarities and differences between our result with the minimal coupling cases of $f(R,\Lm)$ gravity \cite{Harko2010.EurPhysJC.70.373} and $f(R,T)$ gravity \cite{Harko2011.PRD.84.024020}. Section \ref{sec:04} presents the desired modified fluid model for EDE. Conclusions are presented in Sec. \ref{sec:05}.

\section{General theory}\label{sec:02}
We adopt the simplest spacetime dynamics and focus on generalizing perfect fluid. The action takes the form~\footnote{Our conventions: We adopt the SI Units and retain all physical constants. The metric signature is $(-,+,+,+)$. Operators $\nabla_\mu$ and $\p_\mu$ denote covariant and partial derivatives, respectively. The Christoffel symbols, Riemann, Ricci, Einstein tensors, and Ricci scalar are given by
\begin{gather*}
  \Gamma^\lambda_{\mu\nu}=g^{\lambda\alpha}(\p_\nu g_{\mu\alpha}+\p_\mu g_{\nu\alpha}-\p_\alpha g_{\mu\nu})/2, \\
  R^\rho_{\ \lambda\mu\nu}=\p_\mu\Gamma^\rho_{\lambda\nu} - \p_\nu\Gamma^\rho_{\lambda\mu}
                           + \Gamma^\rho_{\alpha\mu}\Gamma^\alpha_{\lambda\nu} - \Gamma^\rho_{\alpha\nu}\Gamma^\alpha_{\lambda\mu}, \\
  R_{\mu\nu}=R^\alpha_{\phantom{\alpha}{\mu\alpha\nu}}, \
  G_{\mu\nu}=R_{\mu\nu}-g_{\mu\nu}R/2, \
  R=R^\mu_{\ \mu},
\end{gather*}
respectively. We define proper time $\tau$ by $\dx s^2=-c^2\dx\tau^2$, so that the four-velocity $u^\mu\equiv\dx x^\mu/\dx\tau$ satisfies $u^\mu u_\mu=-c^2$. In particular, $\rho c^2$ equals to the energy density, which is different with the conventions in \cite{Harko2010.EurPhysJC.70.373} and \cite{Hawking1973.book}, but is consistent with conventional conventions in cosmology.}
\begin{equation}\label{eq:01}
  S=S_\textsc{eh}+S_\textsc{f}=\int\dx^4x\sqrt{-g}\left[\frac{R}{2\kappa}+\Lf\right],
\end{equation}
where $\kappa=8\pi G/c^4$, $g={\rm det}(|g_{\mu\nu}|)$, and $\Lf$ is a general modified fluid Lagrangian. Variation of the Einstein-Hilbert action with respect to the metric gives $\delta S_\textsc{eh}=\int\dx^4x\sqrt{-g}G_{\mu\nu}\delta g^{\mu\nu}/(2\kappa)$ \cite{Misner1973.book}. Variation of the fluid action can be written formally as $\delta S_\textsc{f}=-\int\dx^4x\sqrt{-g}T_{\mu\nu}\delta g^{\mu\nu}/2$, where $T_{\mu\nu}$ is the energy-momentum tensor. These variations give the Einstein field equations $G_{\mu\nu}=\kappa T_{\mu\nu}$, which in turn give $\nabla_\nu T^{\mu\nu}=0$ based on the Bianchi identity.

More properties about the fluid are needed to derive an explicit expression for $T_{\mu\nu}$. We assume that $\Lf$ satisfies $\delta\Lf=(\dx\Lf/\dx n)\delta n$ and the fluid satisfies particle number conservation
\begin{equation}\label{eq:02}
  \nabla_\mu(nu^\mu)=0,
\end{equation}
where $n$ is the particle number density and $u^\mu$ is the four-velocity of the fluid. The first assumption is used to emphasize that no derivative term of $\delta n$ appears in $\delta\Lf$. These two assumptions or their equivalents are widely used to derive the energy-momentum tensors of perfect fluid \cite{Taub1954.PhysRev.94.1468,Hawking1973.book} and beyond \cite{Harko2010.EurPhysJC.70.373,Harko2011.PRD.84.024020}. \citet{Hawking1973.book} present a simple way to derive $\delta n$. They start by rewriting Eq.~(\ref{eq:02}) as $(1/\sqrt{-g})\times\p(\sqrt{-g}nu^\mu)/\p x^\mu=0$, which means $\delta(\sqrt{-g}nu^\mu)=0$. Then the variation of $n^2c^2=g^{-1}(\sqrt{-g}nu^\mu\sqrt{-g}nu^\nu)g_{\mu\nu}$ gives
\begin{equation}\label{eq:03}
  \delta n=\frac{n}{2}(g_{\mu\nu}+\frac{u_\mu u_\nu}{c^2})\delta g^{\mu\nu}.
\end{equation}
Considering the expressions of $\delta\sqrt{-g}$ and $\delta\Lf$, we obtain
\begin{equation}\label{eq:04}
  T_{\mu\nu}=-n\frac{\dx\Lf}{\dx n}\frac{u_\mu u_\nu}{c^2}+(\Lf-n\frac{\dx\Lf}{\dx n})g_{\mu\nu}.
\end{equation}
In principle, how fluid participates in gravitational interactions is determined by $\Lf$. We can directly specify an expression for $\Lf(n)$, such as $\Lf\propto n$. In this case, the fluid participates in gravitational interactions in the form of particle number. Alternatively, we can also assume that $\Lf$ directly depends on other thermodynamic quantities, such as $\Lf\propto\rho$, where $\rho$ is the fluid mass density \cite{Note1}. In this case, the source of the gravitational interaction is $\rho$ and other related quantities, rather than $n$ as in the previous case. As we show later, this case requires an additional equation to determine the dependence of $\rho$ and $n$, and this equation cannot be given by the variational principle. Note that both cases satisfy $\delta\Lf=(\dx\Lf/\dx n)\delta n$ formally. Now we discuss the above two cases around the theoretical self-consistency.

Neglecting the spacetime dynamics, if we specify an explicit expression for $\Lf(n)$, then there are five variables $\{n,u^\mu\}$ to describe the fluid but six evolution or constraint equations $\{\nabla_\nu T^{\mu\nu}=0$, $u^\mu u_\mu=-c^2$, Eq.~(\ref{eq:02})\}. The system is overdetermined as there are more equations than unknowns. However, the system is still self-consistent as these six equations are not independent of each other. To see this, we start from $u_\mu\nabla_\nu T^{\mu\nu}=0$. Substituting Eq.~(\ref{eq:04}) into this equation, we obtain
\begin{align}\label{eq:05}
  0 &= u_\mu\nabla_\nu T^{\mu\nu}=\nabla_\nu(T^{\mu\nu}u_\mu)-T^{\mu\nu}\nabla_\nu u_\mu, \nonumber\\
    &= \nabla_\nu(\Lf u^\nu)-(\Lf-n\frac{\dx\Lf}{\dx n})\nabla_\nu u^\nu, \nonumber\\
    &= u^\nu\nabla_\nu\Lf+n\frac{\dx\Lf}{\dx n}\nabla_\nu u^\nu, \nonumber\\
    &= \frac{\dx\Lf}{\dx n}\nabla_\nu(nu^\nu),
\end{align}
where the second line uses $u^\mu u_\mu=-c^2$ and its derivative $u^\mu\nabla_\nu u_\mu=0$ \cite{Misner1973.book}, and the fourth line uses the chain rule $\nabla_\nu\Lf=(\dx\Lf/\dx n)\nabla_\nu n$. Therefore, Eq.~(\ref{eq:02}) can be derived from $\{\nabla_\nu T^{\mu\nu}=0, u^\mu u_\mu=-c^2\}$. For gravitational theories with fluid models given by explicit $\Lf(n)$, the Einstein field equations together with $u^\mu u_\mu=-c^2$ are complete and self-consistent. Note that, in this case, it is not necessary to introduce other fluid thermodynamic quantities such as mass density $\rho$ and pressure $p$.

However, other thermodynamic quantities, e.g., $\rho$, are needed to describe perfect fluid \cite{Taub1954.PhysRev.94.1468,Hawking1973.book}. If we introduce such a quantity into the fluid Lagrangian, then we have one more variable to describe the fluid. At the same time, we need one more equation to determine the motion of the fluid. This equation cannot be obtained from the gravitational field equations or variational principle. For clarity, here we assume that $\Lf$ is an explicit function of $\rho$, then there are six variables $\{\rho,n,u^\mu\}$ to describe the fluid but only five independent equations $\{\nabla_\nu T^{\mu\nu}=0$, $u^\mu u_\mu=-c^2\}$. Note that one can repeat the proof given by Eq.~(\ref{eq:05}) as long as $\dx\rho/\dx n$ exists. In principle, the additional equation can be arbitrary since the existing equations are underdetermined. In order to be consistent with conventional thermodynamics, we can adopt the local energy conservation equation \cite{Misner1973.book,Hawking1973.book}
\begin{equation}\label{eq:06}
  n\frac{\dx\rho}{\dx n}=\frac{p}{c^2}+\rho,
\end{equation}
where $p=p(\rho)$ is given by the ordinary known equation of state (EOS) of the fluid. Here we only consider the isentropic fluid. This is widely used in the studies of modified fluid \cite{Harko2010.EurPhysJC.70.373,Harko2011.PRD.84.024020,Harko2020.PRD.101.108501}, and is reasonable in many gravitational processes involving fluid, such as big bang nucleosynthesis \cite{Pitrou2018.PhysRep.754.1}, cosmic recombination \cite{Zeldovich1969.SovPhysJETP.28.146}, and neutron star \cite{Oertel2017.RMP.89.015007}. We would like to highlight that $p$ appearing in Eq.~(\ref{eq:06}) is an auxiliary variable to complete the equation, rather than given directly by the variational principle. Adopting Eq.~(\ref{eq:06}) allows us to discard the possible gravitational particle creation process \cite{Harko2014.PRD.90.044067,Pinto2022.PRD.106.044043} in our framework. Note that particle cannot be created in classical field theory, and the creation is a quantum process. We believe that the modified fluid theory is classical, rather than quantum. This is the key reason for our pursuit of energy conservation. For gravitational theories with fluid models given by explicit $\Lf(\rho)$, the equations $\{G_{\mu\nu}=\kappa T_{\mu\nu}$, $u^\mu u_\mu=-c^2$, Eq.~(\ref{eq:06})$\}$ are complete and self-consistent. The above discussion demonstrates our core strategy for modifying fluid theory. More complex fluid Lagrangian will be discussed later and compared with existing methodologies in the literature.

In order to demonstrate the principle discussed above more intuitively, here we present a cosmological application. The Universe is assumed to be described by the flat Friedmann-Lema\^itre-Robertson-Walker metric $\dx s^2=-c^2\dx t^2+a^2\dx\mathbf{x}^2$, where $a=a(t)$, and the four-velocity $u^\mu=(1,0,0,0)$. Substituting these results into the Einstein field equations with Eq. (\ref{eq:04}), we obtain
\begin{subequations}\label{eq:07}
\begin{gather}
  H^2=-\frac{\kappa c^2}{3}\Lf, \label{eq:07a}\\
  \frac{\ddot{a}}{a}=-\frac{\kappa c^2}{3}\left(\Lf-\frac{3n}{2}\frac{\dx\Lf}{\dx n}\right), \label{eq:07b}
\end{gather}
\end{subequations}
where the Hubble parameter $H\equiv\dot{a}/a$ and $\dot{}\equiv\dx/\dx t$. Independent of $\Lf$, Eq.~(\ref{eq:07}) gives $\dot{n}+3Hn=0$, which is exactly Eq.~(\ref{eq:02}). If $\Lf=\Lf(n)$, then Eq.~(\ref{eq:07}) is complete as there are two equations and two variables $\{a,n\}$. If $\Lf=\Lf(\rho)$, then Eq.~(\ref{eq:07}) is not complete as no equation determines the evolution of $\rho$. In this case, one equation such as Eq.~(\ref{eq:06}) is required. For the photon gas contained in the Universe, regardless of the expression of $\Lf(\rho)$, we can adopt Eq.~(\ref{eq:06}) with $p=\rho c^2/3$ so that $n\propto a^{-3}$ and $\rho\propto a^{-4}$. Therefore, such fluid is consistent with conventional thermodynamics.

\section{$\boldsymbol{f(\chi)}$ fluid}\label{sec:03}
Perfect fluid is the main gravitational source in general relativity. Its Lagrangian can be written as $\Lf=-\rho c^2$ \cite{Hawking1973.book,Note1} and the energy-momentum tensor is generally written as $T_{\mu\nu}^\textsc{(pf)}=(\rho+p/c^2)u_\mu u_\nu+pg_{\mu\nu}$ \cite{Tian2020.PRD.101.063531}. We emphasize that all we obtain from the variational principle is Eq.~(\ref{eq:04}). The appearance of $p$ in $T_{\mu\nu}$ is caused by substituting Eq.~(\ref{eq:06}) into Eq.~(\ref{eq:04}) with $\Lf=-\rho c^2$. The essence of $u_\mu\nabla_\nu T^{\mu\nu}=0$ is particle number conservation Eq.~(\ref{eq:02}) as depicted by Eq.~(\ref{eq:05}), rather than energy conservation as widely believed in the literature.

One generalization of the perfect fluid is to write the Lagrangian as $\Lf=f(\chi)$, where $\chi$ is a scalar related to the fluid, e.g., $n$, $\rho$ or the trace of the conventional energy-momentum tensor $T^\textsc{(pf)}\equiv g^{\mu\nu}T_{\mu\nu}^\textsc{(pf)}=3p-\rho c^2$. In our framework, the gravitational field equations of the first two cases have been discussed before, and case $\chi=T^\textsc{(pf)}$ is formally identical to case $\chi=\rho$.

This generalization includes the minimal coupling cases of $f(R,\Lm)$ gravity \cite{Harko2010.EurPhysJC.70.373} and $f(R,T)$ gravity \cite{Harko2011.PRD.84.024020}. Here is a comparison of our results with those given in the literature \cite{Harko2008.PLB.669.376,Harko2010.PRD.81.044021,Harko2010.EurPhysJC.70.373, Minazzoli2012.PRD.86.087502,Wang2012.CQG.29.215016,Harko2011.PRD.84.024020,Sharif2012.JCAP.03.028,Barrientos2018.PRD.97.104041}. In the series of work on $f(R,\Lm)$ gravity \cite{Harko2008.PLB.669.376,Harko2010.PRD.81.044021,Harko2010.EurPhysJC.70.373,Minazzoli2012.PRD.86.087502}, the authors used $\rho$ to denote \textit{rest} mass density \cite{Note1}, and obtained $\delta\rho$ from rest mass conservation. This is essentially the same as our discussion of Eqs.~(\ref{eq:02}) and (\ref{eq:03}). They then analyzed gravitational applications by treating $\Lm$ as an explicit function of $\rho$, which is similar to the case of $\Lf=\Lf(n)$ in our discussions. For the case of minimal coupling between spacetime and matter, they obtained a result similar to our Eq.~(\ref{eq:04}), and then rewrote the result in the form of $T_{\mu\nu}^\textsc{(pf)}$ with redefined mass/energy density and pressure. Finally a given EOS can be used to reconstruct the explicit expression of $\Lm(\rho)$ (see Sec.~II in \cite{Minazzoli2012.PRD.86.087502} for an example). In summary, their result suggests that one $\Lm(\rho)$ corresponds to one specific EOS if the fluid is still perfect. Note that this procedure aims to reconstruct the Lagrangian of perfect fluid, not to generalize the fluid. This is self-consistent, and the result should be equivalent to those given directly in the perfect fluid case. Here we illustrate this equivalence with an example. In our conventions, Eq.~(\ref{eq:04}) and the form of $T_{\mu\nu}^\textsc{(pf)}$ give the redefined mass density $\tilde{\rho}=-\Lf/c^2$ and pressure $\tilde{p}=\Lf-n\dx\Lf/\dx n$. Here the tilde represents redefinition. These redefined quantities satisfy $n\frac{\dx\tilde{\rho}}{\dx n}=\frac{\tilde{p}}{c^2}+\tilde{\rho}$ as $u^\mu\nabla^\nu T_{\mu\nu}^\textsc{(pf)}=0$. If the EOS $w(n)\equiv \tilde{p}/(\tilde{\rho}c^2)$ is known, then $\Lf(n)$ is determined by
\begin{equation}\label{eq:08}
  \frac{n}{\Lf}\frac{\dx\Lf}{\dx n}=w+1.
\end{equation}
For the photon gas ($w=1/3$), the above equation gives $\Lf\propto n^{4/3}$, which is consistent with the result obtained in the conventional perfect fluid framework (see the analysis of the photon gas in an expanding Universe in the perfect fluid framework).

Considering the above discussion and the composition of functions, one might guess that any fluid Lagrangian can be regarded as $\Lf(n)$, so that Eq.~(\ref{eq:01}) can only describe the perfect fluid. In this idea, the physical mass density and pressure should be redefined as discussed above Eq.~(\ref{eq:08}), and the redefined quantities satisfy conventional conservation laws. This is essentially the core of the comment on $f(R,T)$ gravity given by \cite{Fisher2019.PRD.100.064059,Fisher2020.PRD.101.108502}. However, in our opinion, this is not true. In principle, the minimal coupling case of $f(R,T)$ gravity is intend to modify the perfect fluid, rather than reconstruct its Lagrangian. The core of modifying fluid lies in the relationship between $\Lf$ and the physical mass density $\rho$. We can still generalize the perfect fluid by modifying $\Lf(\rho)$ as we discussed earlier. We agree with the reply given by \cite{Harko2020.PRD.101.108501} that the prior $\rho$ has physical thermodynamic interpretation, and the mass density should not be redefined based on a conservation law. In particular, there is a counterexample to \cite{Fisher2019.PRD.100.064059,Fisher2020.PRD.101.108502}. In our framework, both the prior $\rho$ and the redefined $\tilde{\rho}$ formally satisfy the conservation law Eq.~(\ref{eq:06}) even if $\Lf(\rho)$ is general. There is no reason to define the physical mass density by the \textit{latter} one as did in \cite{Fisher2019.PRD.100.064059,Fisher2020.PRD.101.108502}. Compared with the minimal coupling case of $f(R,T)$ gravity \cite{Harko2011.PRD.84.024020}, our theory can naturally incorporate the conservation law Eq.~(\ref{eq:06}), and no gravitational particle creation process \cite{Harko2014.PRD.90.044067,Pinto2022.PRD.106.044043} is required.

\section{EDE in $\boldsymbol{f(\rho,w)}$ fluid}\label{sec:04}
Similar to $f(R,w)$ gravity we mentioned but not analyzed in \cite{Tian2021.PRD.103.043518}, here we use $f(\rho,w)$ fluid to realize the EDE triggered by cosmic radiation-matter transition. We adopt the Lagrangian
\begin{equation}\label{eq:09}
  \Lf=-\rho c^2\times\left[1+\alpha\sin^\beta(3w\pi)\right],
\end{equation}
where the dimensionless parameters $\alpha=\mathcal{O}(0.1)$ and $\beta=\mathcal{O}(1)$, and the conventional fluid EOS $w\equiv p/(\rho c^2)$. For our EDE purpose, the fluid here includes neutrino, photon, baryon and dark matter. The function $\sin(3w\pi)$ is chosen such that the modification vanishes at $w=0$ and $1/3$. The parameters $\alpha$ and $\beta$ control the amplitude and width of $\Omega_\textsc{ede}$, respectively. This realization does not need to specify any energy scale. For the gravitational theory with Eq.~(\ref{eq:09}), the complete and self-consistent field equations are $\{G_{\mu\nu}=\kappa T_{\mu\nu}$ with Eq.~(\ref{eq:04}), $u^\mu u_\mu=-c^2$, Eq.~(\ref{eq:06})$\}$. Note that here $\frac{\dx\Lf}{\dx n}=\frac{\p\Lf}{\p\rho}\frac{\dx\rho}{\dx n}+\frac{\p\Lf}{\p w}\frac{\dx w}{\dx n}$.

\begin{figure}[!t]
  \centering
  \includegraphics[width=1\linewidth]{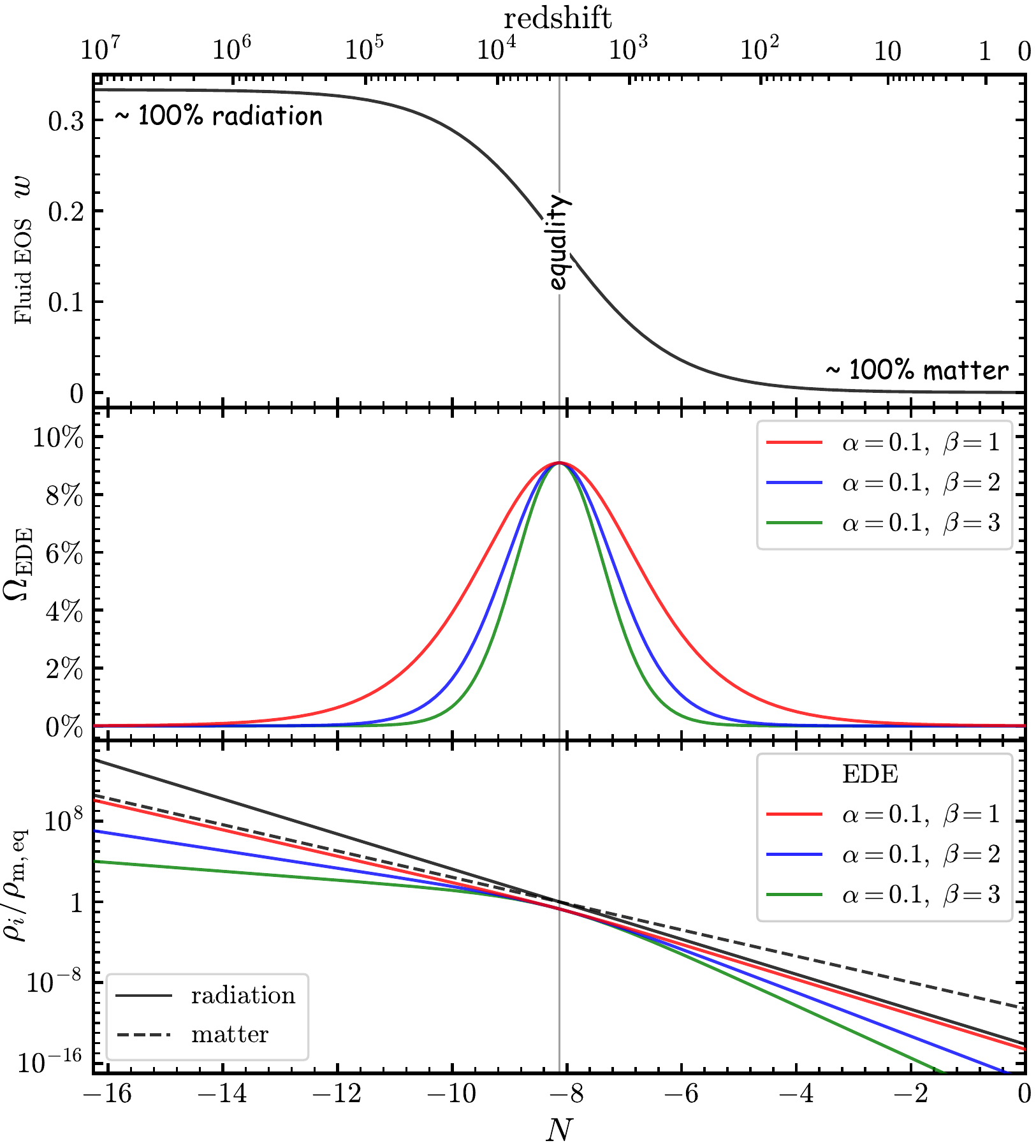}
  \caption{Cosmological evolution of the EDE triggered by cosmic radiation-matter transition and realized in $f(\rho,w)$ fluid. The $\rho_i$ denotes the density of radiation (neutrino and photon, $\propto a^{-4}$), matter (baryon and dark matter, $\propto a^{-3}$) and EDE [$=(\rho_{\rm r}+\rho_{\rm m})\times\alpha\sin^\beta(3w\pi)$], and is rescaled by the matter density at equality $\rho_{\rm m,eq}$. The $\Omega_\textsc{ede}$ and $w$ can be found in the main text. The top axis denotes the cosmological redshift.}
  \label{fig:01}
\end{figure}

For the flat Universe, the complete cosmic evolution equations can be chosen as Eqs.~(\ref{eq:02}), (\ref{eq:06}) and (\ref{eq:07a}). The $w$ is a given variable to characterise the fluid, and Eq.~(\ref{eq:07b}) can be derived from this set of equations. The Friedmann equation~(\ref{eq:07a}) gives the relative energy density of EDE
\begin{equation}
  \Omega_\textsc{ede}=\frac{\alpha\sin^\beta(3w\pi)}{1+\alpha\sin^\beta(3w\pi)}.
\end{equation}
We define the e-folding number $N\equiv\ln(a/a_0)$, where $a_0$ is the cosmic scale factor today. Then $w=(1/3)/[1+\exp(N-N_{\rm eq})]$ for the real Universe contains radiation and pressureless matter \cite{Tian2021.PRD.103.043518}, where $N_{\rm eq}=-8.13$ corresponds to matter-radiation equality \cite{Aghanim2020.AaA.641.A6}. Figure~\ref{fig:01} plots the cosmic evolutions of $w$, $\Omega_\textsc{ede}$ and the density $\rho_i$. The parameter $\alpha=0.1$ roughly corresponds to $\Omega_\textsc{ede}\approx10\%$ at matter-radiation equality, which is the preferred value given by cosmological parameter constraints
\cite{Agrawal2019,Poulin2019.PRL.122.221301,Smith2021.PRD.103.123542,Hill2022.PRD.105.123536,Smith2022.PRD.106.043526}. After the equality, we require EDE dilutes away at least as fast as radiation, which corresponds to $\beta\geq1$ (see the bottom part of Fig. \ref{fig:01}). The model with $\beta\geq1$ also exhibits well in the radiation-dominated era. This figure confirms that Eq.~(\ref{eq:09}) completely realizes the idea that EDE triggered by radiation-matter transition, and solves the relevant coincidence problem.

In the limit of $w\rightarrow0$, we obtain the pressureless perfect fluid from Eq.~(\ref{eq:09}), and then $\nabla_\nu T^{\mu\nu}=0$ gives the geodesic equations $u^\nu\nabla_\nu u^\mu=0$ \cite{Thorne2017.book}. In the solar system, planet can be regarded as pressureless fluid element. Therefore, the planet moves along the geodesic even though the fluid Lagrangian reads Eq.~(\ref{eq:09}). A non-zero $w$ may affect the motion of the stars, e.g., neutron star. This effect may leave an imprint on the gravitational waveforms of binary neutron star mergers. There is another mechanism leading to similar influences. The $w$-modification can affect the structure of neutron star and thus the gravitational waves from binaries through tidal interactions \cite{Flanagan2008.PRD.77.021502,Damour2009.PRD.80.084035,Binnington2009.PRD.80.084018, De2018.PRL.121.091102,Annala2018.PRL.120.172703,Abbott2018.PRL.121.161101,Pratten2022.PRL.129.081102}. These effects may be observable by future gravitational wave detectors with optimum sensitivity range from decihertz \cite{Kawamura2019.IntJModPhysD.28.1845001} to kilohertz \cite{Ackley2020.PubAstronSocAust.37.e047}. Analysis of these issues will be presented in the future.

\section{Conclusions}\label{sec:05}
A general framework to modify perfect fluid is presented in this paper. The proof given by Eq.~(\ref{eq:05}) paves the way of constructing the complete and self-consistent field equations, and allows the modified fluid to satisfy energy conservation. Comparisons between our result and previous work are discussed in detail. Our variational method and result for $T_{\mu\nu}$ are similar to those in \cite{Harko2020.PRD.101.108501}. The difference is that we highlight that Eq. (\ref{eq:06}) needs to be introduced separately, and cannot be given by the variational principle. Our $\Lf(n)$ case is equivalent to the minimal coupling case of $f(R,\Lm)$ gravity \cite{Harko2010.EurPhysJC.70.373}. For the debate on $f(R,T)$ gravity \cite{Harko2011.PRD.84.024020}, our $f(\chi)$ case provides evidence against \cite{Fisher2019.PRD.100.064059,Fisher2020.PRD.101.108502} and supports \cite{Harko2020.PRD.101.108501}, and we conclude that there is no reason to redefine the physical mass density based on the modified fluid Lagrangian or the formally conservation law. Unlike the minimal coupling case of the $f(R,T)$ gravity \cite{Harko2011.PRD.84.024020}, the energy conservation law Eq.~(\ref{eq:06}) can be naturally incorporated in our framework. The nonminimal coupling of spacetime and fluid was not discussed in this paper. This generalization within our framework and a more comprehensive comparison with the $f(R,T)$ gravity will be studied in a future work.

As an application, we propose the $f(\rho,w)$ fluid with Eq.~(\ref{eq:09}) to finish the idea that EDE triggered by radiation-matter transition \cite{Tian2021.PRD.103.043518} --- one way to solve the EDE coincidence problem. There are other ways to address the EDE coincidence, e.g., neutrino-triggered EDE \cite{Sakstein2020.PRL.124.161301,CarrilloGonzalez2021.JCAP.04.063,Gogoi2021.ApJ.915.132,deSouza2023}, dark matter-triggered EDE \cite{Karwal2022.PRD.105.063535,Lin2022.arXiv.2212.08098}, and multiple scaling fields \cite{Sabla2021.PRD.103.103506}. Compared with these models, our model does not require any energy scale, and only introduces two dimensionless parameters of order of $\mathcal{O}(0.1)$ and $\mathcal{O}(1)$. Such property may make the theory more natural.

In the future, gravitational waves from binary neutron star mergers \cite{Flanagan2008.PRD.77.021502,Damour2009.PRD.80.084035,Binnington2009.PRD.80.084018, De2018.PRL.121.091102,Annala2018.PRL.120.172703,Abbott2018.PRL.121.161101,Pratten2022.PRL.129.081102} may be able to provide a cross-check for our EDE model. The possible positive results given by the relevant cross-checking can lead to robust statements about the existence of the $w$-modification of perfect fluid.

\section*{Acknowledgements}
This work was supported by the National Natural Science Foundation of China under Grants No. 12021003, No. 11920101003 and No. 11633001, and the Strategic Priority Research Program of the Chinese Academy of Sciences under Grant No. XDB23000000. S. X. T. was also supported by the Initiative Postdocs Supporting Program under Grant No. BX20200065 and China Postdoctoral Science Foundation under Grant No. 2021M700481.

\end{document}